\documentclass[twocolumn,showpacs,preprintnumbers,amsmath,amssymb,prl]{revtex4-1}
\usepackage{graphicx}
\usepackage{dcolumn}
\usepackage{bm}
\usepackage{color}

\preprint{APS/pre-print}

\begin{document}

\title{A nonlinear elastic instability in channel flows at low Reynolds numbers}

\author{L. Pan$^1$}
\author{A. Morozov$^2$}
\author{C. Wagner$^3$}
\author{P. E. Arratia$^1$}



\affiliation{$^1$Department of Mechanical Engineering and Applied Mechanics, University of Pennsylvania, Philadelphia, USA}
\affiliation{$^2$SUPA, School of Physics and Astronomy, University of Edinburgh, Edinburgh, UK}
\affiliation{$^3$Experimental Physics, Saarland University, Saarbrucken, DE}

\date{\today}

\begin{abstract}

It is presently believed that flows of viscoelastic polymer solutions in geometries such as a straight pipe or channel are linearly stable. Here we present experimental evidence that such flows can be nonlinearly unstable and can exhibit a subcritical bifurcation. Velocimetry measurements are performed in a long, straight micro-channel; flow disturbances are introduced at the entrance of the channel system by placing a variable number of obstacles.  Above a critical flow rate and a critical size of the perturbation, a sudden onset of large velocity fluctuations indicates presence of a nonlinear subcritical instability. Together with the previous observations of hydrodynamic instabilities in curved geometries, our results suggest that any flow of polymer solutions becomes unstable at sufficiently high flow rates.

\end{abstract}


\maketitle

Solutions containing polymer molecules do not flow like water. Even when flowing slowly, these fluids can exhibit hydrodynamic instabilities~\cite{Larson1990JFM, Shaqfeh1996ARFM, Larson1992RA, Arratia2006PRL, Groisman2003S, Poole2007PRL, McKinley1993, Pakdel1996PRL} and a new type of turbulence - the so-called \emph{purely elastic turbulence}~\cite{Groisman2000N, Groisman2004NJP} even at low Reynolds numbers (Re). These phenomena, driven by the anisotropic elasticity of the fluid, were experimentally observed only in geometries with sufficient curvature, like rotational flows between two cylinders \cite{Larson1990JFM, Groisman1998PF, Muller1989RA} and plates \cite{McKinley1996JNNFM}, in curved channels \cite{Groisman2001N, Groisman2004NJP}, and around obstacles \cite{Arora2002JNNFM}. Most of the nonlinear flow behavior observed in these studies arises from the extra elastic stresses due to the presence of polymer molecules in the fluid. These elastic stresses are history dependent and evolve on the time-scale $\lambda$ that in dilute solutions is proportional to the time needed for a polymer molecule to relax to its equilibrium state \cite{Larson1999}.

A common feature of the above-mentioned geometries is the presence of curved streamlines in the base flow with a sufficient velocity gradient across the streamlines. It has been argued that this is a necessary condition for infinitesimal perturbations to be amplified by the normal stress imbalances in viscoelastic flows~\cite{Larson1990JFM,Pakdel1996PRL,McKinley1996JNNFM}. This condition can be written as $(\lambda\,U N_{1})/(R\,\Sigma)\geq M$~\cite{Pakdel1996PRL, McKinley1996JNNFM,Morozov2007}, where $M$ is a constant that only depends on the type of flow geometry, $U$ is a typical velocity along the streamlines, $R$ is the radius of streamline curvature, and $N_{1}$ and $\Sigma$ are the first normal stress difference and the shear stress, correspondingly. According to this condition,
purely elastic \emph{linear} instabilities are not possible when the curvature of the flow geometry is zero, and infinitesimal perturbations decay at a rate proportional to 1/$\lambda$~\cite{Pakdel1996PRL, Ho1977JNNFM, Wilson1999}.

\begin{figure}[tbhp!]
  \includegraphics[width=6.0cm]{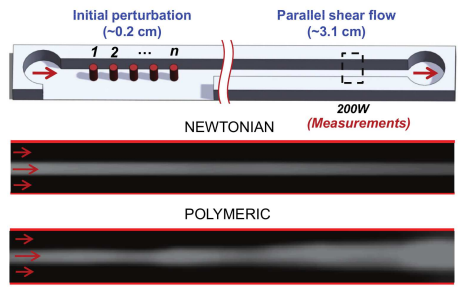}
  \caption{Color Online. (a) Sketch of the experimental setup. The dash line window represents a typical sampling position. (b, c) Sample snapshots of dye advection experiments at $\emph{Re}<0.01$ and 15 cylinders: (b) Newtonian case; (c) polymeric case, \emph{Wi}=10.9. Field of view is 9$W$.}
\end{figure}

Nevertheless, the absence of a linear instability does not imply absolute stability. Indeed, recent theoretical \cite{Meulenbroek, Morozov2005PRL, Morozov2007, Hoda2008JFM, Jovanovic2010PF} and indirect experimental \cite{Bertola2003PRL, Bonn2011PRE} evidence points towards a finite-amplitude transition in viscoelastic flows with parallel streamlines even at low Reynolds number, where viscous and elastic forces dominate over inertial forces. An earlier study of the flow of a polymeric melt extruded out of a thin cylindrical capillary~\cite{Bertola2003PRL} reported that outside the capillary the extrudate developed periodic surface modulations above a critical flow rate. While this behavior was shown to be hysteretic, and thus consistent with a nonlinear instability scenario, it was unclear whether the instability originated inside or outside of the capillary. In a recent investigation of a dilute polymer solution flowing in a straight pipe~\cite{Bonn2011PRE}, the authors observed unusually large velocity fluctuations inside the pipe, but the subcritical nature of the instability was not established and no hysteric behavior was reported. Here, we present the first direct experimental evidence of a nonlinear subcritical instability in a wall-bounded straight channel flow for a single-phase viscoelastic fluid.

Experiments are performed in a long ($\sim$ 3.3 cm), straight microchannel system that consists of a short initial perturbation region ($\sim$ 0.2 cm) followed by a long parallel flow region ($\sim$ 3.1 cm) as shown in Fig. 1a. The microchannel system is 90~$\mu$m deep and 100~$\mu$m wide. The initial perturbation region is located at the very beginning of the channel and contains an array of variable number of cylinders. This short array is only responsible for introducing flow perturbations into the long parallel flow region. Cylinders in the array are 50~$\mu$m in diameter and 90~$\mu$m tall; the distance between two adjacent cylinders is 200~$\mu$m (center to center). The number of cylinders \emph{n} in the initial perturbation region varies from 1 to 15 in order to alter the strength of the perturbation; a channel devoid of cylinders \emph{n}=0 is also used for control. We note that the parallel flow region is long ($\sim$ 3.1 cm), straight, and devoid of cylinders; the fate of the initial flow perturbations introduced by the cylinders is monitored in this long parallel shear flow region using dye advection and velocimetry methods. All microchannels are fabricated using standard soft-lithography methods~\cite{McDonald2002}.

Both Newtonian and polymeric fluids are investigated. The Newtonian fluid is a 90\% by weight glycerol aqueous solution with shear viscosity $\eta \approx$ 0.2 Pa$\cdot$s. The polymeric solution is made by adding 300 ppm of polyacrylamide (PAA, 18 x 10$^{6}$ MW) to a viscous Newtonian solvent (90\% by weight glycerol aqueous solution). Both fluids are characterized using a strain-controlled rheometer at 23 $^{\circ}$C~\cite{SM}. For all experiments, the Reynolds number (Re) is small ($<$0.01) due to the channel small length scale and high fluid viscosity. Here Re=$\rho UL/\eta$ , where \emph{U} is the fluid velocity, \emph{L} is a characteristic length scale, and $\rho$ is the fluid density. The magnitude of the elastic stresses compared to viscous stresses is characterized by the Weissenberg number~\cite{Magda1993, Bird} defined as \emph{Wi}=\emph{N$_1$}/(2$\dot{\gamma}$$\eta$), where \emph{N$_1$} is the first normal stress difference and $\dot{\gamma}$ is the shear-rate; further details in~\cite{SM}.

We begin with dye advection experiments which are performed by injecting small amounts of dyed fluid (fluorescein) into the flow from the top wall using a multilayer injection scheme. Dyed fluid is injected at approximately 1.0~cm downstream from the initial perturbation region in order to display only the flow patterns in the parallel flow region.  Images are taken about 1 mm downstream from injection point. Figures 1(b) and (c) show snapshots of the dye advection experiments at Re$<10^{-2}$ for both the Newtonian and polymeric cases, respectively for a channel containing 15 cylinders ($n=15$). The Newtonian case (Fig. 1b) shows a stable layer of dyed fluid that does not mix with the undyed fluid except by diffusion. An entirely different pattern is observed when the Newtonian fluid is replaced by a polymeric solution at $Wi=10.9$ (Fig. 1c). The dyed fluid quickly mixes with the undyed fluid, which suggests the presence of hydrodynamic instabilities and time-dependent flow. Below we show that this time-dependent flow is not due to the downstream advection of the fluctuations around the cylinders, but rather is a unique non-linear state independent of the original perturbation.

\begin{figure}[t]
  \includegraphics[width=8.0cm]{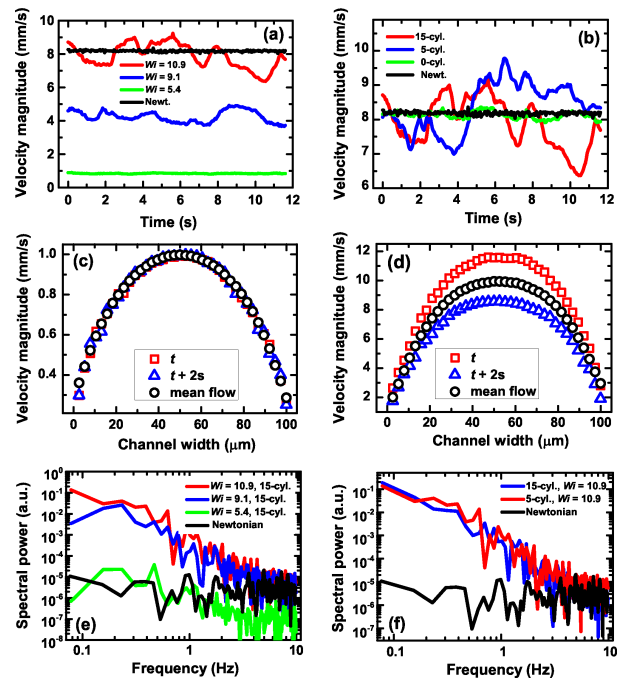}
  \caption{Color Online. (a,b): Spatially averaged velocity magnitude as a function of time for (a) \emph{n}=15 as a function of \emph{Wi} and for (b) \emph{Wi}=10.9 as a function of \emph{n}. (c,d): Instantaneous and mean streamwise velocity profiles for \emph{n}=15 for (c) \emph{Wi} = 5.4 and (d) \emph{Wi} = 10.9. (e,f): Power spectra of the velocity fluctuations from (a) and (b).}
\end{figure}

Particle velocimetry methods are used to quantify the instability observed in the dye experiments. The flow is seeded with small fluorescent particles (0.86~$\mu$m in diameter) that are tracked using a CMOS (3 kHz) and an epi-fluorescent microscope. The particle tracks are measured at a mid-point between the top and bottom plates of the channel in order to minimize the effects of out-of-plane velocity gradients; the thickness of the measuring plane is approximately 2 $\mu$m. Measurements are performed in several locations along the channel including one channel width (1W) after the last cylinder as well as 50W, 100W, 150W, and 200W. The 1W measuring location is used to monitor the amplitude of the initial disturbance introduced in the flow by the array of \emph{n} cylinders. The other measuring locations are used to monitor the fate of the initial disturbance in the parallel flow region.

To quantify the time dependence of the flow, we sample a square area (about 35\% of the channel width centered at midpoint) of the velocity fields in the parallel shear flow region, and measure the average streamwise speed as a function of time. The sampling rates are long enough ($\sim$~1 ms) to ensure the accuracy of the velocimetry measurement but are much shorter than the typical time scale of the fluid motion. All measurements shown in Fig. 2 are taken at 200W or 2~cm downstream from the last cylinder. Figure 2(a) shows samples of the velocity magnitude records measured far downstream (200W) for a channel with \emph{n}=15 as function of \emph{Wi} or equivalently flow rate. We find that, for the polymeric case, the velocity fluctuations become larger as \emph{Wi} is increased. The Newtonian case, on the other hand, produces no such fluctuations at comparable shear-rates (210 s$^{-1}$). In Fig. 2(b), we show velocity records of the polymeric solution at a fixed \emph{Wi}=10.9 for channel systems with different number of cylinders \emph{n}. For the case of an empty channel (\emph{n}=0), the viscoelastic case shows no significant fluctuations even at the highest shear-rate. Time-dependent velocity fluctuations, however, become apparent as cylinders are introduced in the channel. This is further illustrated by plotting the instantaneous and mean velocity profiles of the polymeric fluid. For \emph{Wi}=5.4 (Fig. 2c), there is no time-dependence and the (base) flow is unidirectional. For  \emph{Wi}=10.9 (Fig. 2d), on the other hand, the instantaneous velocity profiles show significant differences between each other and with the mean profile. Importantly, the amplitude of velocity fluctuations is roughly independent of \emph{n} at a fixed \emph{Wi} (Fig. 2b). 

The corresponding power spectra of the velocity signals in Fig. 2(a) and (b) are shown in Fig. 2(e) and (f), respectively. Since the entire velocity field must be measured at each instant, the records are only a few hundred points long, but this is sufficient to establish the qualitative features of the spectra. In Fig. 2(e), we note that the spectral power at low frequencies grows by 2-4 orders of magnitude as the \emph{Wi} is increased at a fixed \emph{n}. Similar behavior is observed as \emph{n} is increased at a fixed \emph{Wi}. The velocity fluctuations are non-periodic, with a possible power-law decay, indicating that the flow is excited at many time-scales. Such decay has been observed in many flow geometries with curved streamlines and has been interpreted as evidence of elastic turbulence~\cite{Groisman2000N, Groisman2004NJP}. By contrast, the Newtonian and \emph{n}=0 (polymeric) cases show a relatively flat power spectra consistent with noise.

\begin{figure}[t]
  \includegraphics[width=5.5cm]{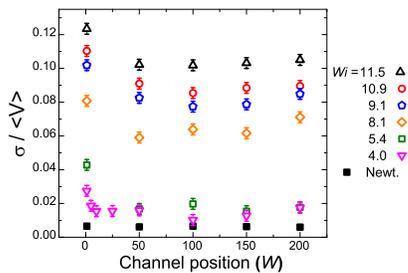}
  \caption{Color Online. Velocity fluctuations $\sigma$/$<$\emph{V}$>$ along the parallel flow region as a function of channel position and \emph{Wi}. }
 \end{figure}

Next, we investigate how the amplitude of the velocity fluctuations changes along the channel. In Fig. 3, we plot the standard deviation $\sigma$ of the velocity signal normalized by its mean $<$\emph{V}$>$ measured at different locations in the parallel shear region as function of \emph{Wi} for \emph{n}=15. As expected, the velocity fluctuations for the Newtonian case are small ($\sim$ 0.01) and independent of the channel position even at high shear-rates. Results for the polymeric solution show a different behavior. For \emph{Wi} up to 5.4, the values of $\sigma$/$<$\emph{V}$>$ are relatively large immediately after the last cylinder (1\emph{W}) due to a well-known instability that develops in the wake of a cylinder for viscoelastic flows \cite{McKinley1993, James1970JFM}. In our experiments this instability sets in at \emph{Wi}$\approx$3.5. However, the velocity fluctuations decay to values close to the Newtonian case in just a few channel widths. This indicates that any flow disturbance that initially develops in the channel is short-lived and damped by viscous forces, and the flow far downstream is stable for $\emph{Wi}<5.4$.

An entirely different behavior emerges for $\emph{Wi}>5.4$ and \emph{n}=15. The velocity fluctuations, created in the wake of the array of cylinders, settle to values of $\sigma$/$<$\emph{V}$>$ that are significantly larger than the Newtonian case (Fig. 3). These velocity fluctuations decay to only 8-10\% after 50\emph{W}, and remain approximately constant thereafter even at 200\emph{W}. Similar behavior in $\sigma$/$<$\emph{V}$>$ is observed down to $\emph{Wi}>8.1$. This data strongly suggests that a time-dependent flow can be created and sustained in the parallel shear flow region provided the Weissenberg number and the strength of the initial perturbation supplied by the flow around cylinders are both sufficiently large.

\begin{figure}[t]
  \includegraphics[width=7.5cm]{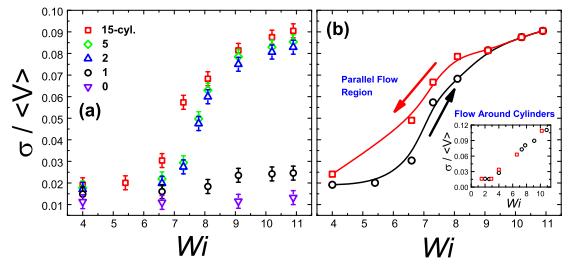}
  \caption{Color Online. Velocity fluctuations $\sigma/<$\emph{V}$>$ at 200\emph{W} downstream from the last cylinder:  (a) as a function of \emph{Wi} and \emph{n}; (b) as a function of \emph{Wi} for \emph{n}=15 as the shear-rate is increased (dark curve) and decreased (red curve). Lines are added to guide the eye. (Inset) Bifurcation diagram for the instability around the cylinders.}
\end{figure}

We now study how large a perturbation (created by the cylinders) should be to destabilize the flow in the parallel shear region. In Fig. 4(a), the magnitude of the velocity fluctuations far downstream from the last cylinder are plotted as a function of \emph{Wi} for channels with different number of cylinders \emph{n}. For \emph{n}=0, no instability is found anywhere in the channel, and the values of $\sigma$/$<$\emph{V}$>$ remain near 1\%. For \emph{n}=1, a relatively small levels of fluctuations are observed ($\sim$ 2.5\%) even for large \emph{Wi}; these fluctuations could be due to flow convection.

A notable difference in flow behavior is observed when the number of obstacles is further increased. For \emph{n}=2 and $\emph{Wi}<5.4$, the values of $\sigma$/$<$\emph{V}$>$ are still relatively small ($\sim$ 2\%). However, for $\emph{Wi}>5.4$, the velocity fluctuations sharply increase and reach an asymptotic value of about 9\% for large \emph{Wi}. Similar behavior to the \emph{n}=2 case is observed for the \emph{n}=5 and \emph{n}=15 cases in polymer solutions. The data in Fig 4(a) clearly shows the development of two branches after a critical value of \emph{Wi}, one in which the flow is stable ($\emph{n}<2$) and the other in which the flow is unstable (n$\geq$2). Importantly, for $n\geq$2 the level of fluctuations saturates and does not depend on \emph{n} suggesting that the flow has reached the same nonlinear state independent of the initial perturbation. A phase diagram showing this nonlinear instability is available in~\cite{SM}.

The dynamic behavior of the flow transitions are also investigated (Fig. 4b). For the linear array of cylinders (Fig. 4b, inset), where measurements are performed immediately after the last cylinder or at 1W, the transition from steady to unsteady flow is characterized by a forward bifurcation and no hysteric behavior \cite{McKinley1993}.  On the other hand, the transition from steady to unsteady flow in the parallel shear region (or 200W) exhibits a dynamical hysteresis (Fig. 4b): upon the increase or decrease of the flow rate, the level of fluctuations sharply rises and falls at different \emph{Wi}'s. This hysteric behavior is a hallmark of a subcritical bifurcation. This ultimately proves that, while the cylinders or obstacles play an important role in providing strong initial perturbations to the flow, the resulting bulk instability in the parallel shear flow section of the channel is clearly distinct from the instability around the cylinders.

Finally, we comment on the relation between the observed transition and the non-normal growth theory developed for Newtonian \cite{Trefethen1993,SchmidBook} and viscoelastic \cite{Doering2006, Hoda2008JFM, Jovanovic2010PF} shear flows. This theory considers {\it linear} dynamics of perturbations and predicts that in a linearly stable system a perturbation which is not a pure eigenmode of the non-normal linear operator will grow algebraically in time before decaying exponentially. Our observations are, however, incompatible with this prediction. 
Indeed, such time-evolution in a frame co-moving with the mean flow translates into an initial spatial region of increasing fluctuations followed by a region where fluctuations decay, when viewed in the lab frame. Instead, above the transition we observe fluctuation levels that are essentially independent of the spatial position downstream of the channel (Fig. 3). Moreover, the non-normal growth theory predicts that {\it any} non-modal perturbation would be amplified. But no significant fluctuations far downstream of the channel are found in the presence of one cylinder even after the linear instability around the cylinder sets in (Fig. 4a).  This suggests that there is a finite-amplitude threshold for the transition and that the transition is non-linear. Finally, the subcritical nature of the transition, demonstrated in Fig.4 (b), rules out any explanation based on a linear theory.  While it is likely that the non-normal growth is a part of the transition we observed, the transition itself is a nonlinear phenomenon, similar to the Newtonian case \cite{Waleffe1997,Schoppa2002}.


In summary, we have shown experimentally the existence of a nonlinear subcritical instability for polymeric fluids in a parallel shear flow at low Re. The critical value of \emph{Wi} for the onset of the subcritical instability in the parallel flow is larger than 5.2 for the type of disturbances introduced here. This critical value may, however, be very sensitive to the type and strength of the initial perturbation and will be further investigated. A possible mechanism leading to this subcritical instability has been proposed \cite{Morozov2005PRL} in which the initial finite amplitude disturbance produces a new effective base flow with curved streamlines in the parallel flow region and becomes linearly unstable \cite{Groisman2003S, Pakdel1996PRL}. The transition then would depend on whether the disturbance is sufficiently strong and long-lived to become unstable. This scenario is akin to the transition to turbulence of Newtonian fluids in pipe and channel flows, except that the instability is caused by the nonlinear elastic stresses and not inertia \cite{Hof2004S, Avila2011S}.


We thank W. van Saarloos, R. Poole, A. Lindner, and D. Bartolo for discussions and comments on the manuscript. We also thank R. Sureshkumar, M. Graham, R. Larson, B. Khomami, S. Kumar and M. Shatz for fruitful discussions and X. Shen for assistance with rheology. AM acknowledges support from the EPSRC Career Acceleration Fellowship (grant EP/I004262/1). PEA acknowledges support from NSF-Career CBET-0954084.


%

\end{document}